\documentclass[]{iopart}

\usepackage[T1]{fontenc}
\usepackage{lipsum}  
\usepackage{graphicx}
\usepackage[english]{babel}
\usepackage{upgreek}

\expandafter\let\csname equation*\endcsname\relax

\expandafter\let\csname endequation*\endcsname\relax

\usepackage{amsmath}
\usepackage{amssymb}
\usepackage{tabls}
\usepackage{dsfont}
\usepackage[colorlinks=true,citecolor=blue,linkcolor=red,urlcolor=blue]{hyperref}
\usepackage{cite}
\usepackage{citesort}
\usepackage{bbm}
\usepackage{braket} 

\makeatletter
\def\bbl@set@language#1{%
	\edef\languagename{%
		\ifnum\escapechar=\expandafter`\string#1\@empty
		\else\string#1\@empty\fi}%
	\@ifundefined{babel@language@alias@\languagename}{}{%
		\edef\languagename{\@nameuse{babel@language@alias@\languagename}}%
	}%
	\select@language{\languagename}%
	\expandafter\ifx\csname date\languagename\endcsname\relax\else
	\if@filesw
	\protected@write\@auxout{}{\string\select@language{\languagename}}%
	\bbl@for\bbl@tempa\BabelContentsFiles{%
		\addtocontents{\bbl@tempa}{\xstring\select@language{\languagename}}}%
	\bbl@usehooks{write}{}%
	\fi
	\fi}
\newcommand{\DeclareLanguageAlias}[2]{%
	\global\@namedef{babel@language@alias@#1}{#2}%
}
\makeatother

\newcommand\varpm{\mathbin{\vcenter{\hbox{%
  \oalign{\hfil$\scriptstyle+$\hfil\cr
          \noalign{\kern-.3ex}
          $\scriptscriptstyle({-})$\cr}%
}}}}
\newcommand\varmp{\mathbin{\vcenter{\hbox{%
  \oalign{$\scriptstyle({+})$\cr
          \noalign{\kern-.3ex}
          \hfil$\scriptscriptstyle-$\hfil\cr}%
}}}}

\DeclareLanguageAlias{en}{english}

\begin{document}

\title{Quantum Gates with Oscillating Exchange Interaction}

\author{Daniel Q. L. Nguyen, Irina Heinz and Guido Burkard}
\address{Department of Physics, University of Konstanz, D-78457 Konstanz, Germany}
\eads{\mailto{daniel.nguyen@uni-konstanz.de}, \mailto{irina.heinz@uni-konstanz.de}, \mailto{guido.burkard@uni-konstanz.de}}



\begin{abstract}
Two-qubit gates between spin qubits are often performed using a rectangular or an adiabatic exchange interaction pulse resulting in a CZ gate. An oscillating exchange pulse not only performs a CZ gate, but also enables the iSWAP gate, which offers more flexibility to perform quantum algorithms. We provide a detailed description for two-qubit gates using resonant and off-resonant exchange pulses, give conditions for performing the respective gates, and compare their  performance to the state-of-the-art static counterpart. We find that for relatively low charge noise the gates still perform reliably and compare to the conventional CZ gate.
\end{abstract}

\maketitle

\section{Introduction}
In recent years advances in silicon and germanium based quantum dot spin qubits \cite{Loss_1998, Zwanenburg_2013,Burkard2023} showed immense potential as a platform to realize quantum computers \cite{Xue2022,Noiri2022,M_dzik_2022}. In particular highly enriched silicon reduces the interaction of an electron spin qubit with nuclear spins, and thus enables long coherence times and high-fidelity qubit operations. Single-qubit gates are enabled by electric dipole spin resonance (EDSR), where an oscillating gate voltage causes electron modulation in a magnetic gradient field and with it an effective magnetic driving field on the qubit. The exchange interaction between neighboring electron spins is electrically controlled via a barrier gate voltage to perform two-qubit gates \cite{PhysRevB.59.2070, Yoneda_2017, Watson_2018, Zajac_2017, PhysRevLett.107.146801}. Charge noise is often suppressed to first order by operating at a symmetric operation point (``sweet spot'')~\cite{PhysRevLett.116.116801, PhysRevLett.116.110402, PhysRevLett.115.096801}, while dephasing effects can be reduced through a large energy splitting due to a strong magnetic field gradient~\cite{nichol2016highfidelity} realized by a micromagnet \cite{Yoneda_2015, Kawakami_2014}. 

So far high-fidelity two-qubit gates, as the CNOT or CZ, were realized using a constant exchange interaction between two electron spins \cite{Xue2022, Noiri2022}. However it was already experimentally demonstrated \cite{Sigillito_2019}, that an oscillating exchange interaction enables the iSWAP gate which constitutes an alternative two-qubit gate for universal quantum computation. In this paper we present a theoretical treatment of two-qubit gates generated by an oscillating  exchange interaction between two neighboring spins.  We derive conditions for the realization of various gates and investigate their performances compared to the conventional case with a static exchange coupling.

This paper is organized as follows. In Section \ref{sec:model} we introduce the system Hamiltonian and the Makhlin invariants, which we will use in the reminder of this work. In Section \ref{sec:results} we summarize our results found for the resonant case (Sec. \ref{subsec:resonant-case}) and non-resonant case (Sec. \ref{subsec:non-resonant-case}). Finally, Section \ref{sec:influence-of-noise} estimates the effect of noise in experimental setups on the respective gates and compares performances between oscillating and static exchange interactions.

\begin{figure}
	\centering
	\includegraphics[width=0.45\textwidth]{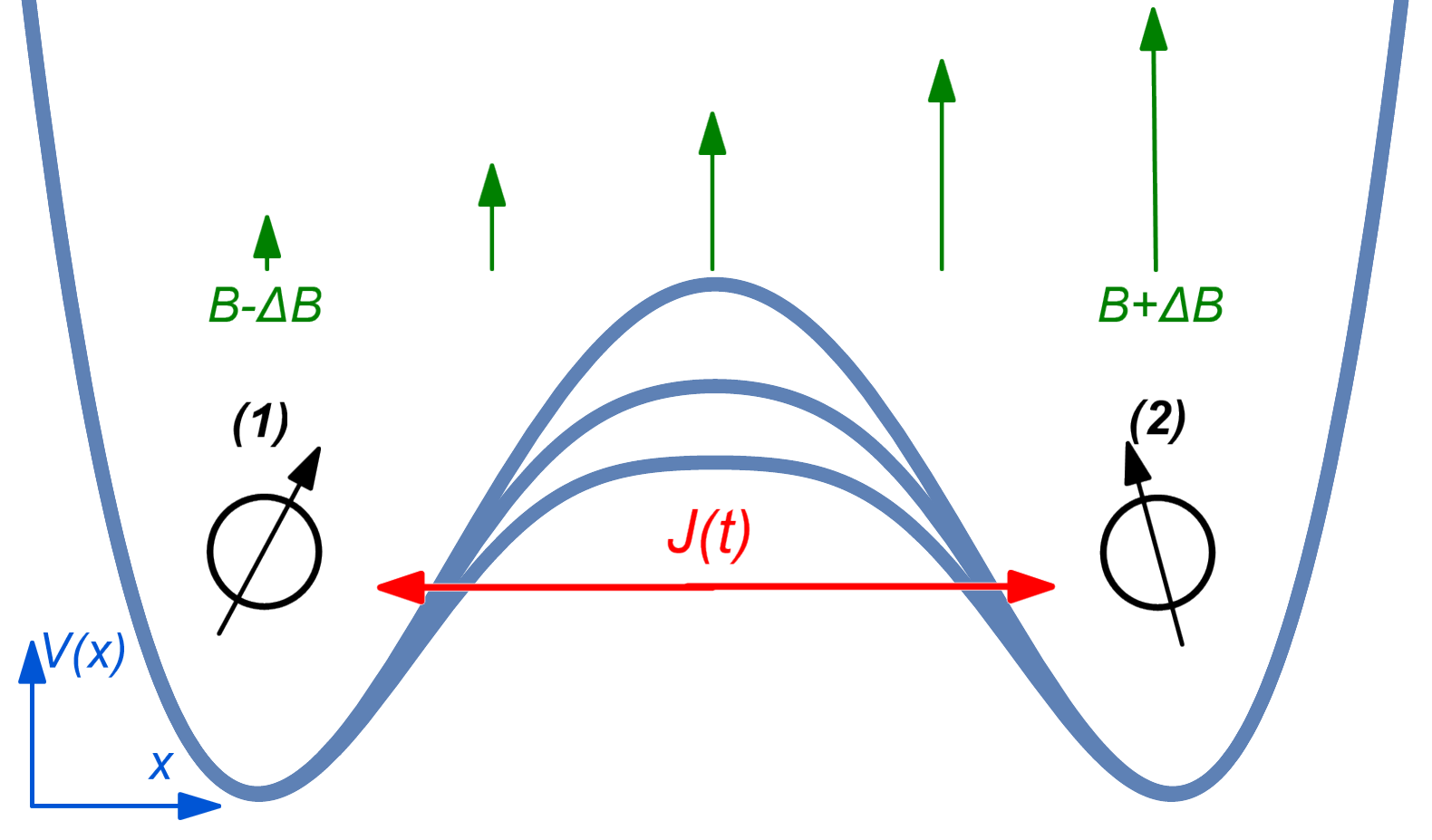}
	\caption{Schematic model of the two-qubit system. Two electron spins are localized by plunger and barrier gate potentials. The magnetic gradient field $B\pm\Delta B$ enables individual addressing of the single spins. Controlling the middle barrier voltage $V(x)$ between the electrons can realize a time dependent exchange interaction $J(t)$ between the electron spins 1 and 2.}
	\label{Fig:Schematic2QubitSystem}
\end{figure}

\section{Model and Methods\label{sec:model}}
We consider a two-qubit system realized by two electrons located in two coupled quantum dots, as shown in figure~\ref{Fig:Schematic2QubitSystem}. It is possible to control the exchange interaction $J(t)$ between the electrons by tuning the middle barrier between them. Additionally, there is a magnetic field parallel to the $z$-axis with $B-\Delta B$ at site 1 and $B+\Delta B$ at site 2. The Hamiltonian describing the local Zeeman splitting and the exchange coupling between the electrons reads as follows,
\begin{align}\label{eq:HamiltonianWithExchange}
    \begin{aligned}
    H=&B\left( \hat{S}_{1z}+\hat{S}_{2z} \right)+\Delta B \left(\hat{S}_{2z}-\hat{S}_{1z} \right) +J(t)\left(\mathbf{\hat{S}}_1\cdot \mathbf{\hat{S}}_2-\frac{1}{4}\mathbbm{1}\right)
    = H_1+H_2,
    \end{aligned}
\end{align}
where $H_1=B(\hat{S}_{1z}+\hat{S}_{2z})$, 
and $\mathbf{\hat{S}}_{1}$ and $\mathbf{\hat{S}}_{2}$ refer to the spin operators of the electron at site 1 and 2, and where we consider an oscillating exchange interaction $J(t)=J_0+J_1\cos\left(\omega t\right)$. In case of a static or adiabatic exchange interaction $J_1=0$, the Hamiltonian is transformed into a rotating frame and can result in a CZ gate when choosing parameters and timing accordingly \cite{Russ_2018}, where $|J|\ll |\Delta B|$.

A useful tool to find and identify two-qubit gates from a non-trivial time evolution $U$ are the Makhlin invariants $G_1$ and $G_2$, defined as $G_1=\text{Tr}[M(U)]^2 \det [U]/16$ and $G_2=(\text{Tr}[M(U)]^2-\text{Tr}[M(U)^2])\det [U]/4$ with $M(U) = U_B^T U_B$ and $U_B$ is the transformed time evolution $U$ in the Bell basis \cite{Makhlin2002}. In particular, two two-qubit gates $A,B\in \text{SU}(4)$ differ only by single-qubit operations $U_1\otimes U_2, V_1\otimes V_2\in \text{SU}(2)^{\otimes 2}$, i.e.,
$
	A=(V_1\otimes V_2) B (U_1\otimes U_2)
$,
if and only if their Makhlin invariants $G_1$ and $G_2$ are identical.
The relevant Makhlin invariants for this work are $(G_1,G_2)=(1,3)$, $(0,1)$, $(0,-1)$, for the identity, CZ, and iSWAP gates. 

As a quantitative measure, we calculate the fidelity between the desired ideal gate  $U_\text{ideal}$ and the actual gate $U_\text{actual}$ using \cite{Pedersen_2007}
\begin{align}
F=F\left(U_\text{ideal},U_\text{actual}\right)=\frac{d+\left|\text{Tr}\left[U_\text{ideal}^\dagger U_\text{actual}\right]\right|^2}{d\left(d+1\right)},
\end{align}
where $d$ denotes the dimension of the Hilbert space. Here, $d=4$ for two qubits.

\section{Results\label{sec:results}}
Here, we consider an oscillating exchange interaction $J(t)=J_0+J_1\cos\left(\omega t\right)$ between the two electron spins, where we restrict ourselves to $0 \leq |J_1| \leq J_0$ where the total exchange is positive $J\ge 0$. In this case one can find approximate analytical solutions to the time evolution of the Hamiltonian~\eqref{eq:HamiltonianWithExchange} for either
   (A) a resonant frequency $\omega=2\Delta B$,
   or
(B) a far-detuned drive, $\left|2\Delta B-\omega\right|\gg J_1$.

\subsection{Resonant case} \label{subsec:resonant-case}
First, we set $\omega=2\Delta B$. By separating the Hamiltonian~\eqref{eq:HamiltonianWithExchange} into two commuting parts $H=H_1+H_2$,
it is possible to construct the time evolution $U=U_1 U_2$ generated by $H$ by calculating the time evolution $U_1$ and $U_2$ of $H_1$ and $H_2$ respectively. 
The part $H_1$ describing the homogeneous magnetic field $B$ is a time-independent diagonal matrix. The operator $H_2$ is nonzero only in the subspace spanned by the basis vectors $\{ \ket{\uparrow\downarrow}, \ket{\downarrow\uparrow} \}$. By reducing $H_2$ to a single-qubit Hamiltonian $H_2\rightarrow h_2$ in the basis $\{ \ket{\uparrow\downarrow}, \ket{\downarrow\uparrow} \}$ it is again possible to separate $h_2$ into two commuting parts $h_2=h_{21}+h_{22} $ with
\begin{align}
    h_{21}&=-\frac{1}{2}J(t)\mathbbm{1} ,\\
    h_{22}&=-\Delta B \sigma_z + \frac{1}{2}J(t)\sigma_x , 
\end{align}
where $\sigma_{x,z}$ denote the Pauli matrices. $h_{21}$ is a diagonal matrix for which the analytical solution to its time evolution $u_{21}$ is known. For $h_{22}$ one can use the rotating wave approximation (RWA) which first requires the use of a rotating frame $r=\exp\left(-i\frac{\omega}{2}\sigma_z t\right)$ resulting in the unitary transformation
$h_{22}\rightarrow i\Dot{r}r^{-1}+rh_{22}r^{-1}=\tilde{h}_{22}$.
As a condition for the RWA it is required that $\left|\frac{J_0}{\omega}\right|,\left|\frac{J_1}{\omega}\right|\ll 1$ such that fast oscillating terms can be neglected and
\begin{align}
	\tilde{h}_{22}\approx
 \frac{J_1}{4}\sigma_x
\end{align}
is time independent. Therefore, it is simple to calculate its time evolution $\tilde{u}_{22}=\exp(-it\tilde h_{22})$. By rotating back to the local reference frame $u_{22}=r^{-1}\tilde{u}_{22}$ and writing the time evolution $u_{21}$ and $u_{22}$ as $U_{21}$ and $U_{22}$ respectively, $U$ can be expressed as a product $U\approx U_1 U_{21} U_{22}$.
The resulting approximated time evolution can be written as a matrix in the basis 
$\ket{\uparrow\uparrow},
\ket{\uparrow\downarrow},\ket{\downarrow\uparrow},\ket{\downarrow\downarrow}$,
\begin{align}\label{eq:U-res Solution}
U_\text{res}=\begin{pmatrix}
e^{-iBt} & 0 & 0 & 0 \\
0 & e^{ia_+(t)}\cos(\frac{J_1}{4}t) & -ie^{ia_+(t)}\sin(\frac{J_1}{4}t) & 0 \\
0 & -ie^{ia_-(t)}\sin(\frac{J_1}{4}t) & e^{ia_-(t)}\cos(\frac{J_1}{4}t) & 0 \\
0 & 0 & 0 & e^{iBt} 
\end{pmatrix} ,
\end{align}
where we defined
\begin{align}
a_\pm(t)&:=\frac{1}{2}\left(\left(J_0\pm\omega\right)t+\frac{J_1}{\omega}\sin\left(\omega t\right)\right). \label{Eq:apm}
\end{align}

Using Makhlin invariants we can now give conditions for a set of parameters $(J_0,J_1,t)$ for which the approximated time evolution is locally equivalent to a CZ or iSWAP gate. The Makhlin invariants are given by
\begin{align}
    G_1(U_\text{res})&=\frac{1}{4}\alpha\left(1+\alpha^{-1}\beta\right)^2 ,\\
    G_2(U_\text{res})&=\frac{1}{2}\alpha^{-1}+\frac{1}{2}\alpha+2\beta,
\end{align}
with
\begin{align}
		\alpha&=\exp\left(-i\left(J_0t+\frac{J_1}{\omega}\sin\left(\omega t\right)\right)\right) , \hspace{0.5cm} \beta =\cos\left(\frac{J_1}{2}t\right) .
\end{align}
When choosing parameter set $(J_0,J_1,\tau_{n,m})$ such that
\begin{align}\label{eq:Res:Conditions For Gate Times1}
		J_0\tau_{n,m}+\frac{2m\pi}{\omega \tau_{n,m}}\sin\left(\omega \tau_{n,m}\right)&=n\pi ,\\
    J_1&=\frac{2m\pi}{\tau_{n,m}}\label{eq:Res:Conditions For Gate Times2} ,
\end{align}
where $n,m\in \mathbb{Z}$ we obtain $\alpha, \beta=\pm 1$ depending on the parity of $n$ and $m$. If $n$ is odd and $m$ is even the Makhlin invariants of the approximated time evolution match with  the Makhlin invariants of a CZ gate. Moreover,  for $n$ is even and $m$ is odd the time evolution is instead locally equivalent to an iSWAP gate.
For a given $J_0$ we always find a solution, however, here we additionally require preferably small $n$ for short gate times.

If a set of parameters $(J_0,J_1,\tau_{n,m})$ fulfills conditions~\eqref{eq:Res:Conditions For Gate Times1} and \eqref{eq:Res:Conditions For Gate Times2} with $n=1+2n_1$ and $m=2m_1$ for $n_1,m_1\in\mathbb{Z}$, the resulting time evolution operator is locally equivalent to the CZ gate as previously explained. Depending on the parity of $n_1+m_1$ the approximated time evolution is
\begin{align}\label{eq:Matrixform CZ+-res}
\text{CZ}_\pm^\text{res}=\text{diag}\left(e^{-iB\tau},\pm ie^{\frac{i}{2}\omega \tau},\pm ie^{-\frac{i}{2}\omega \tau},e^{iB\tau}\right)
\end{align}
with $\tau=\tau_{n,m}$, where one obtains $\text{CZ}_+^\text{res}$ if $n_1+m_1$ is even and $\text{CZ}_-^\text{res}$ if $n_1+m_1$ is odd.

Similar results are found in the case of a locally equivalent iSWAP gate where we find two gates $\text{iSWAP}_\pm$ both locally equivalent to the iSWAP gate,
\begin{align}\label{eq:Matrixform iSWAP+-}
    \text{iSWAP}_\pm=\begin{pmatrix}
e^{-iB\tau} & 0 & 0 & 0 \\
0 & 0 & \pm ie^{\frac{i}{2}\omega \tau} & 0 \\
0 & \pm ie^{-\frac{i}{2}\omega \tau} & 0 & 0 \\
0 & 0 & 0 & e^{iB\tau} 
\end{pmatrix}\,.
\end{align}
Here $\tau=\tau_{n,m}$ and we obtain $\text{iSWAP}_+$ and $\text{iSWAP}_-$ for $n_2+m_2$ is odd and $n_2+m_2$ is even respectively if $n=2n_2$ and $m=1+m_2$ ($n_2,m_2\in\mathbb{Z}$).

The required single-qubit operations to obtain the CZ and iSWAP gates are $z$ rotations on the two qubits and depend on the values for $B,\omega$ and $\tau_{n,m}$ as well as the values for $n$ and $m$ in Eq.~\eqref{eq:Res:Conditions For Gate Times1} and Eq.~\eqref{eq:Res:Conditions For Gate Times2}. It can be noted that only $z$-rotations applied to the qubits are necessary to transform the locally equivalent CZ gate to an actual CZ gate or the locally equivalent iSWAP gate to an iSWAP gate.

\begin{figure}[hb]
    \centering
    \includegraphics[width=0.48\textwidth]{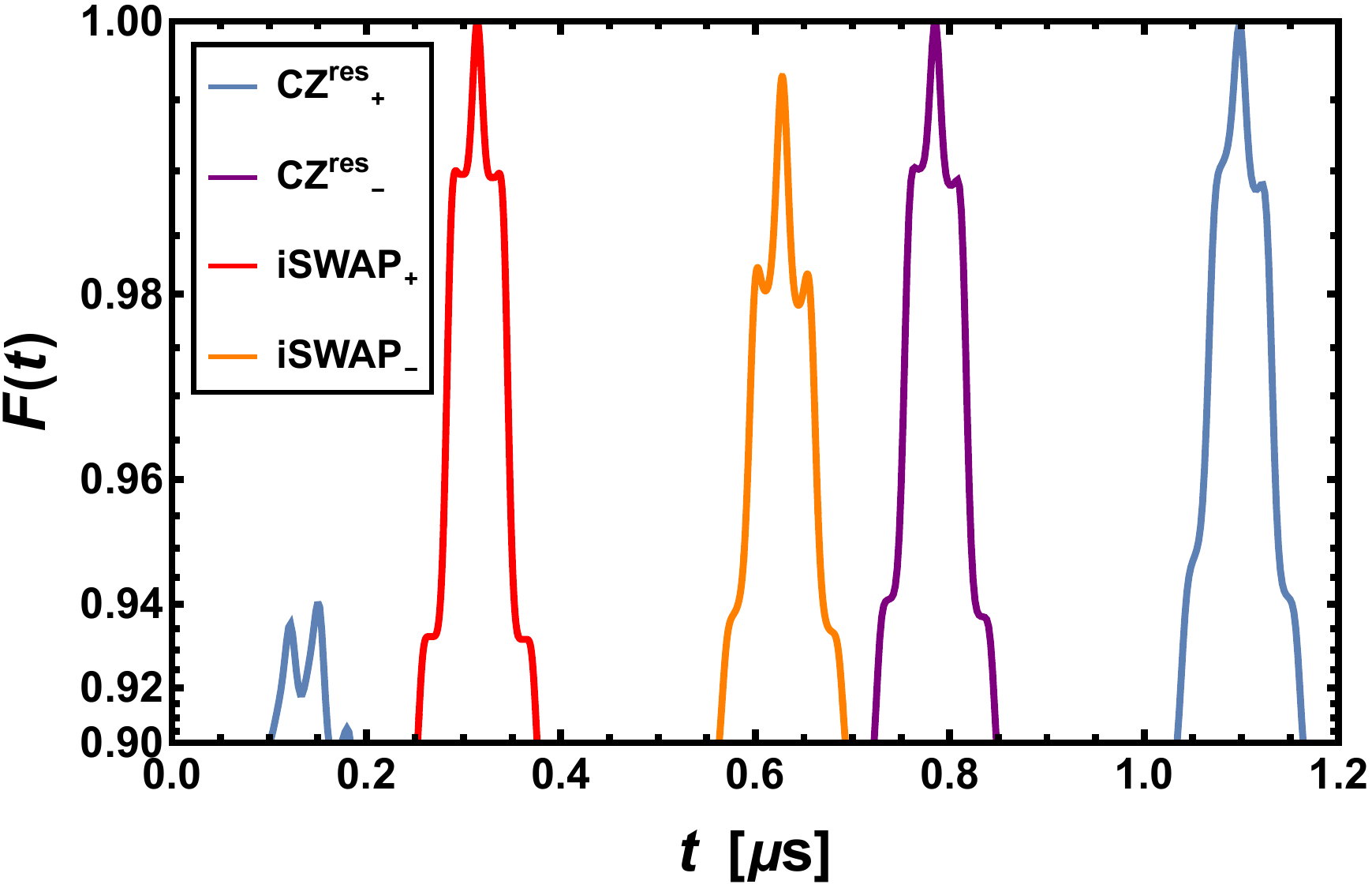}
    \caption{Fidelity $F(U_\text{}(t),X(t))$ between the actual numerical solution and the locally equivalent CZ/iSWAP gates $X(t)=\text{CZ}_\pm^\text{res} (t), \text{iSWAP}_\pm (t)$ as a function of time. The used parameters are $B=1\,\text{GHz},\Delta B=100\,\text{MHz},J_{0}=20\,\text{MHz}, \omega=200\,\text{MHz}$. For $J_1$ we use $11.43\,\text{MHz},16\,\text{MHz},20\,\text{MHz},10\,\text{MHz}$ for the $\text{CZ}_+^\text{res},\text{CZ}_-^\text{res},\text{iSWAP}_+,\text{iSWAP}_-$ comparison, respectively. Numerically we find the gate times $\tau\approx 1.098\,\mu\text{s}, 0.784\,\mu\text{s}, 0.314\,\mu\text{s}, 0.628\,\mu\text{s}$ for these gates.}
    \label{fig:ResFid-tdep}
\end{figure}

Figure~\ref{fig:ResFid-tdep} shows the fidelity over time for a suitable set of parameters for the $\text{CZ}_{\pm}^{\text{res}}$ and $\text{iSWAP}_{\pm}$. The locally equivalent CZ and iSWAP gates are treated as a function of $t$ as they depend on the gate time $\tau_{n,m}$ (see Eq.~\eqref{eq:Matrixform CZ+-res} and Eq.~\eqref{eq:Matrixform iSWAP+-}). With $J_0=20\,\text{MHz}$ a solution to the conditions~\eqref{eq:Res:Conditions For Gate Times1} and \eqref{eq:Res:Conditions For Gate Times2} with $J_1\leq J_0$ requires small $m$. For the shortest gate times the smallest possible $n$ for each gate $\text{CZ}_\pm^\text{res}$ and $\text{iSWAP}_\pm$ are chosen.  Note that the fidelity $F$ uses the actual solution of the time evolution with numerical methods. The gate times of the numerical result in figure \ref{fig:ResFid-tdep} match the times from the conditions stated in Eq.~\eqref{eq:Res:Conditions For Gate Times1} and Eq.~\eqref{eq:Res:Conditions For Gate Times2}. We find that the $\text{iSWAP}_-$ gate has a noticeably lower fidelity than the $\text{iSWAP}_+$ gate. Further calculations show that a lower value for $J_1$ correlates to a lower fidelity of any of the $\text{iSWAP}_\pm$ gates. Thus, the best possible $\text{iSWAP}_\pm$ gate would therefore ideally require $J_1=J_0$.

\subsection{Non-resonant case} \label{subsec:non-resonant-case}
For the far off-resonant case we transform our Hamiltonian~\eqref{eq:HamiltonianWithExchange} in a the rotating frame $H\longrightarrow\Dot{R}R^\dagger+RHR^\dagger =\tilde{H}$ with 
$R=\exp[i(B (\hat{S}_{1z} + \hat{S}_{2z})t]$, and apply the rotating wave approximation with the following conditions,
\begin{align}\label{eq:NonRes-ConditionForRWA}
	\left|\frac{J_0}{4\Delta B}\right|,\left|\frac{J_1}{8\Delta B+4\omega}\right|,\left|\frac{J_1}{8\Delta B-4\omega}\right|&\ll1\,.
\end{align}
This yields an approximated Hamiltonian
$\tilde{H}\approx \text{diag}\left(0,-\Delta B -\frac{1}{2}J(t),\Delta B -\frac{1}{2}J(t),0\right)$,
for which the time evolution is a diagonal matrix
\begin{align}
    \tilde{U}_\text{non-res}= \text{diag}\left(1,\exp\left(i\frac{A_{+}(t)}{2}\right),\exp\left(i\frac{A_{-}(t)}{2}\right),1\right),
    \label{eq:non-res}
\end{align}
with $A_{\pm}(t)=\pm 2\Delta B +J_0t+\frac{J_1}{\omega}\sin\left(\omega t\right)$. We note that the conditions stated in Eq.~\eqref{eq:NonRes-ConditionForRWA} is fulfilled if $\Delta B \gg J_0,J_1$ in combination with the far-off-resonant condition $\left|2\Delta B -\omega\right|\gg 1$. 

The Makhlin invariants of the approximated time evolution are then
\begin{align}
    G_1\left(\tilde{U}_\text{non-res}\right)&=\cos\left(\frac{J_0}{2}t+\frac{J_1}{2\omega}\sin(\omega t)\right)^2 , \label{Eq:non-res-G1}\\
    G_2\left(\tilde{U}_\text{non-res}\right)&=2+\cos\left(J_0t+\frac{J_1}{\omega}\sin(\omega t)\right)  \label{Eq:non-res-G2}.
\end{align}
One can observe that there exist solutions $(J_0,J_1,t)$ such that $\tilde{U}_\text{non-res}$ is locally equivalent to a CZ gate. For this, the set of parameters $(J_0,J_1,\tau_n)$ must fulfill
\begin{align}\label{eq:NonRes:Conditions For Gate Times}
	    J_0\tau_n+\frac{J_1}{\omega}\sin\left(\omega \tau_n\right)\stackrel{!}{=}\pi+2n\pi, \,\qquad n\in\mathbb{N}_0\,.
\end{align}
For given $J_0>0,J_1$, and $n$, a solution for $\tau_n$ always exists. The approximated time evolution with the gate time $\tau_n$ is then as follows,
\begin{align}
    \tilde{U}_\text{non-res}(\tau_n)=
    \text{diag}\left(1, \pm i, \pm i,1\right),
\end{align}
up to single-qubit rotations around the $z$-axis.
The necessary single-qubit operations to transform the time evolution to a CZ gate only depend $\tau_n$. We obtain two different locally equivalent CZ gates to which we will refer as $\text{CZ}_\pm^\text{nres}=\text{diag}\left(1,\pm i,\pm i, 1\right)$.
It should be noted that for this approximation and rotating frame a locally equivalent iSWAP gate is not possible.

For off-resonant drives with $\left|\frac{J_0}{\omega}\right|,\left|\frac{J_1}{\omega}\right|\ll 1$ we can approximate the time evolution more accurately. In the rotating frame $R=\exp[iB (\hat{S}_{1z} + \hat{S}_{2z})t]$ we find the commuting subspace Hamiltonian $h_2=h_{21}+h_{22} $, spanned by the basis vectors $\{ \ket{\uparrow\downarrow}, \ket{\downarrow\uparrow} \}$,
with $h_{21}=-\frac{1}{2}J(t)\mathbbm{1}$, $h_{22}=-\Delta B \sigma_z + \frac{1}{2}J(t)\sigma_x$. Transforming into the rotating frame $h_{22}\rightarrow i\Dot{r}r^{-1}+rh_{22}r^{-1}=\tilde{h}_{22}$ with $r=\exp\left(-i\frac{\omega}{2}\sigma_z t\right)$, analogous to Sec.~\ref{subsec:resonant-case}, we can approximate $\tilde{h}_{22}\approx \frac{J_1}{4}\sigma_x + (\omega/2 - \Delta B) \sigma_z$ and calculate the time evolution. After transforming back to the frame rotating with $R=\exp[iB (\hat{S}_{1z} + \hat{S}_{2z})t]$ we obtain the total time evolution. For a driving time $\tau$ with $\sqrt{J_1^2 + 4(\omega-2\Delta B)^2} \tau /4 = m \pi$ the time evolution becomes diagonal
\begin{align}
    \tilde{U}_{\rm non-res} (\tau) = \text{diag}\left(1, \pm e^{ia_+(\tau)}, \pm e^{ia_-(\tau)},1\right),
\end{align}
with $a_{\pm}$ as in Eq.~\eqref{Eq:apm} and the sign $+$ ($-$) for $m$ even (odd). We then obtain the Makhlin invariants as in Eqs.~\eqref{Eq:non-res-G1} and \eqref{Eq:non-res-G2}, and thus, again find Eq.~\eqref{eq:NonRes:Conditions For Gate Times} as conditions for a CZ gate.

\begin{figure}
	\centering
	\includegraphics[width=0.48\textwidth]{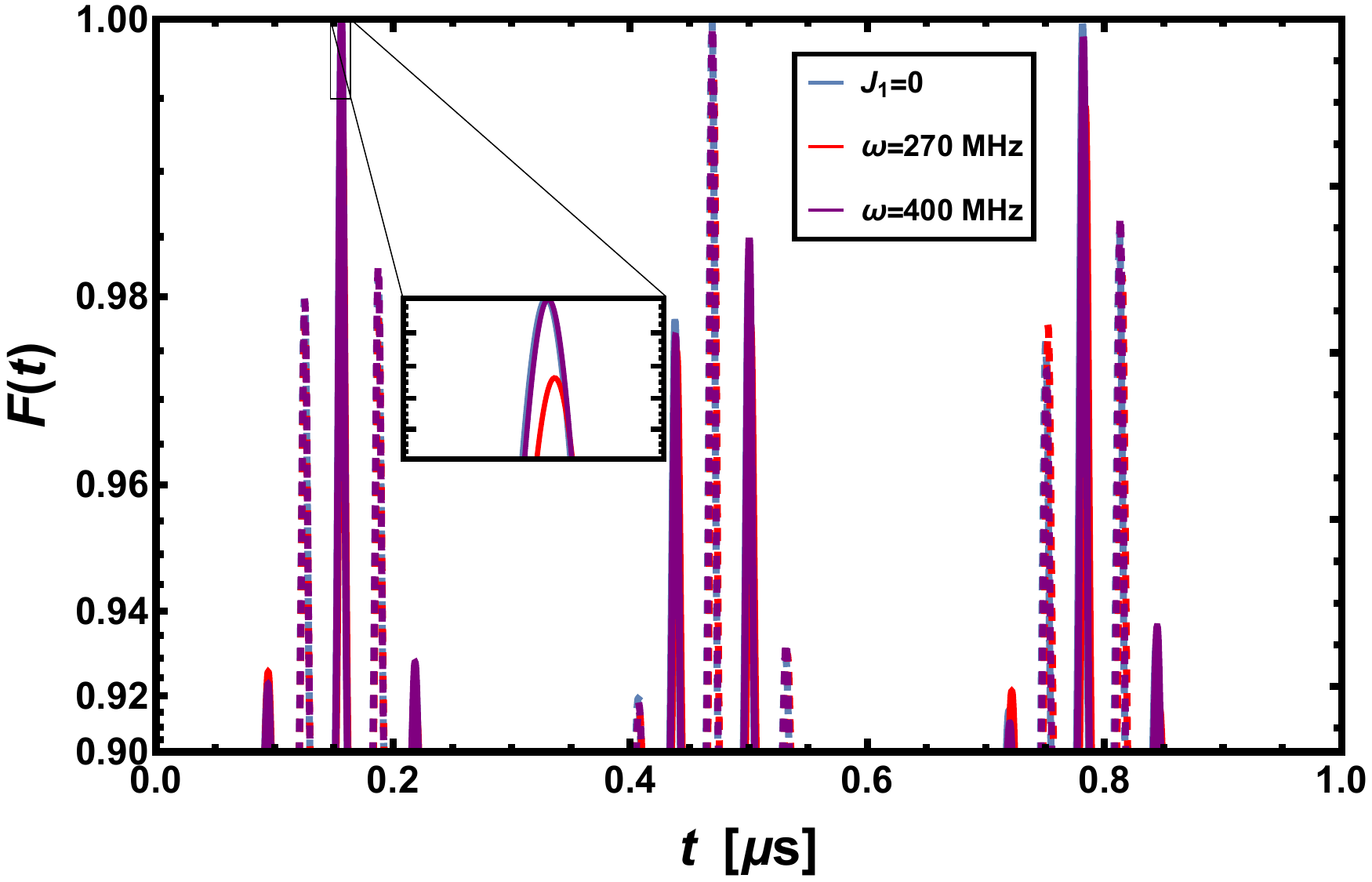}
	\caption{Numerical fidelity $F(\tilde{U}(t),\text{CZ}_\pm^\text{nres})$ as a function of time between the actual solution $\tilde{U}(t)$ in the rotating frame and the locally equivalent CZ gate $\text{CZ}_\pm^\text{nres}$ at different frequencies. Solid (dashed) lines refer to $\text{CZ}_-^\text{nres}$ ($\text{CZ}_+^\text{nres}$). The parameters are chosen as $B=1\,\text{GHz}$, $\Delta B=100\,\text{MHz}$,  and $J_{0}=J_{1}=20\,\text{MHz}$. The case of a constant exchange $J_1=0$ is also shown.}
	\label{Fig:NonResFid-tdep}
\end{figure}

Figure~\ref{Fig:NonResFid-tdep} illustrates the numerical fidelity $F(\tilde{U},\text{CZ}_\pm^\text{nres})$ as a function of time using two different frequencies $\omega=270$ MHz and 400 MHz showing the first three gate times. For comparison, the fidelity for a constant exchange $J_1=0$ is also shown. The fidelity at the maxima corresponds to the first few gate times $\tau_0$, $\tau_1$ and $\tau_2$ and decreases over time. It is therefore better to use the first gate time $\tau_0$. Using an oscillating exchange interaction we find a similar time evolution as in the case of a constant exchange interaction ($J_1=0$). 

We note here that if going into a different rotating frame, one needs to take into account the respective $z$ rotations. Without compensation of these $z$-rotations in the rotating frame $R=\exp[i((B-\Delta B) \hat{S}_{1z} + (B+\Delta B)\hat{S}_{2z})t]$ we obtain a slightly higher fidelity for the oscillating exchange compared to the constant case. The cause of this slight fidelity improvement are off-resonant terms in the Hamiltonian that were neglected in the RWA and lie beyond our simple description. To analyze these higher-order corrections, one could apply a Floquet-Magnus expansion~\cite{Bukov_2015,Blanes_2009,Moore_1990,Mostafazadeh_1997}.

\section{Influence of noise} \label{sec:influence-of-noise}
So far we have considered ideal conditions for constructing two-qubit gates. In experiments under real conditions, however, charge noise in the potential barrier $V(x)$ (see figure~\ref{Fig:Schematic2QubitSystem}) limits the fidelity of two-qubit gates and may impact the results presented in this paper. To model the performance of the derived protocols we introduce charge noise in the exchange coupling. Here we assume the fluctuations in $V$ to be small such that the exchange coupling $J_0(V)$ is linear in $V$.

We estimate the influence of noise in $V$, and therefore $J_0$, to have a Gaussian distribution around the desired value. The influence of noise on the fidelity can then be calculated with
\begin{align}\label{eq:Noise Integral}
    F_\text{N}(\tau,X(\tau))&=\int_\mathbb{R}F(U'(\tau,J_0+J),X(\tau))p(J)dJ ,
\end{align}
where
\begin{align}
    p(J)&=\frac{1}{\sqrt{2\pi}\sigma}\text{exp}\left(-\frac{J^2}{2\sigma^2}\right) ,\\
    X(\tau)&=\text{CZ}_\pm^\text{(n)res}(\tau),\text{iSWAP}_\pm(\tau),
\end{align}
and $U'=U,\tilde{U}$ refers to the actual solutions of the time evolution with and without the rotating frame as required for the resonant and non-resonant case, respectively. The standard deviation $\sigma$ of the Gaussian distribution functions as a measure for varying noise levels. 
For comparison, the noisy fidelity $F_\text{N}$ of the CZ gate with a constant exchange is given by
\begin{equation}\label{eq:NoiseFidelityAtGateTimes}
    F_\text{N}(\tau,X)=\frac{1}{5}\left(3+2e^{-\frac{1}{8}\sigma^2\tau^2}\right),
\end{equation}
with $X=\text{CZ}^\text{const}$ and the gate time $\tau=\tau_n$ fulfills condition~\eqref{eq:NonRes:Conditions For Gate Times} using the analytical solution \eqref{eq:non-res} with $J_1=0$ to the time evolution.

\subsection{Resonant case}\label{sec:noise-resonant}
Using time evolution~\eqref{eq:U-res Solution} for the approximated solution $U_\text{res}$ one can find an analytical solution to the fidelity~\eqref{eq:Noise Integral}. For the locally equivalent CZ (iSWAP) gates the fidelity with noise takes the same form as given in Eq.~\eqref{eq:NoiseFidelityAtGateTimes}. Here we have $X= X(\tau_{n,m})=\text{CZ}_\pm^\text{res}(\tau_{n,m})$ ($\text{iSWAP}_\pm(\tau_{n,m})$)
at the respective gate times $\tau=\tau_{n,m}$ fulfilling the conditions  \eqref{eq:Res:Conditions For Gate Times1} and \eqref{eq:Res:Conditions For Gate Times2}. The decay for $F_\text{N}(\tau_{n,m},X(\tau_{n,m}))$ follows a Gaussian curve but does not, however, fall below a value of $0.6$ for any $n,m$ and $\sigma$.

For the set of parameters used for figure~\ref{fig:ResFid-tdep} one achievable gate time for each $\text{CZ}_-^\text{res}$ (purple curve) and $\text{iSWAP}_+$ (red curve) is visible for the first microsecond. These gate times can be calculated through Eqs.~\eqref{eq:Res:Conditions For Gate Times1} and \eqref{eq:Res:Conditions For Gate Times2} and are in fact the shortest gate times with $J_0=20\,\text{MHz}$ that also fulfill the condition $|J_1|\leq J_0$. With these parameters the integral for $F_\text{N}(\tau_{n,m})$ can be calculated numerically for the different gates.

\begin{figure}
    \centering
    \includegraphics[width=0.48\textwidth]{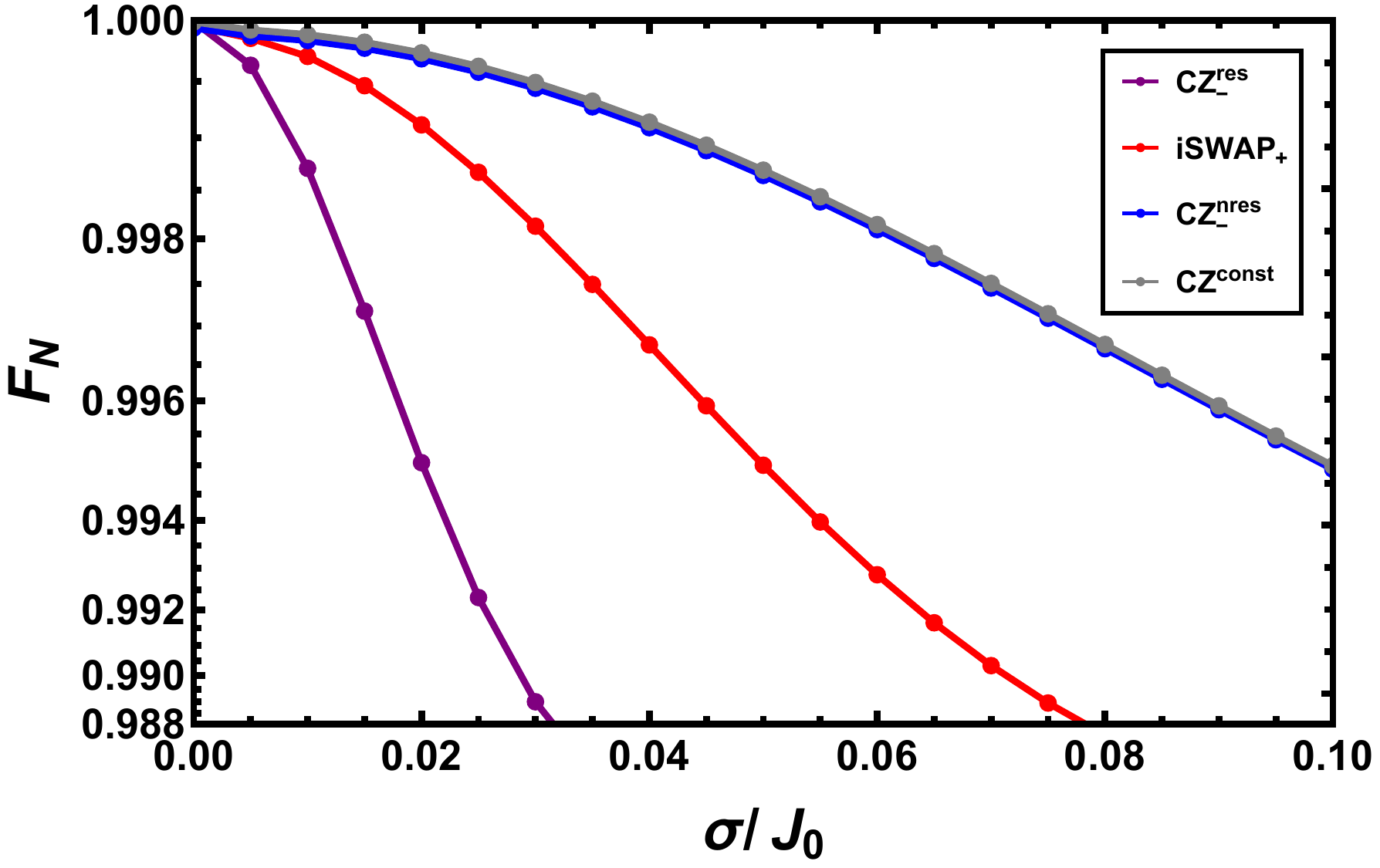}
    \caption{Fidelity $F_\text{N}$ depending on $\sigma/J_0$ for $\text{CZ}_-^\text{res}$, $\text{iSWAP}_+$, $\text{CZ}_-^\text{nres}$  at the gate times determined by Eqs.~\eqref{eq:Res:Conditions For Gate Times1}, \eqref{eq:Res:Conditions For Gate Times2}, and \eqref{eq:NonRes:Conditions For Gate Times}. The case of a constant exchange with gate time $\tau_0=\pi / J_0$ is also shown in gray for reference. The parameters are $B=1\,\text{GHz}$, $\Delta B=-100\,\text{MHz}$,  $J_{0}=20\,\text{MHz}$, and $\omega=200\,\text{MHz}$. We use $J_1=16\,\text{MHz}$ and $20\,\text{MHz}$ for the  $\text{CZ}_-^\text{res}$ and $\text{iSWAP}_+$ gates. The gate times are  $\tau\approx  \pi/4 \,\mu\text{s}, \pi/10\,\mu\text{s}, 0.1563\,\mu\text{s}, 0.1565\,\mu\text{s}$ for the $\text{CZ}_-^\text{res},\text{iSWAP}_+,\text{CZ}_-^\text{nres},\text{CZ}^\text{const}$ fidelity, respectively.}
    \label{fig:FidNoise}
\end{figure}

Figure~\ref{fig:FidNoise} shows the fidelity $F_\text{N}$ at the gate times of the respective quantum gates in the resonant case depending on the noise to signal ratio $\sigma / J_0$. Apparently, shorter gate times benefit the fidelity under influence of noise which the analytical approximation predicts. Furthermore, the case with constant exchange is shown in gray. The two-qubit gates we obtain from an oscillating exchange with a resonant frequency have a similar fidelity as the CZ gate achieved with a constant exchange for rather low charge noise  $\sigma/J_0 \lesssim 1\%$. The constant exchange maintains a high fidelity even for larger noise levels which can be explained by the shorter gate time $\tau$ and the fidelity with noise decaying exponentially in $\tau$.
Note that we find the shortest gate times to be $\tau_{5,2}$ and $\tau_{2,1}$ for the $\text{CZ}^\text{res}$ and iSWAP gate respectively.

\subsection{Non-resonant case}
Similar to the previous section the fidelity affected by noise can be calculated analytically with the approximation $\tilde{U}_\text{non-res}$ giving the same decay as in Eq.~\eqref{eq:NoiseFidelityAtGateTimes} with $X=\text{CZ}_\pm^\text{res}$
at the gate times $\tau_n$ determined by Eq.~\eqref{eq:NonRes:Conditions For Gate Times}.

Using the same parameters as  in figure~\ref{Fig:NonResFid-tdep} one can calculate the first gate time $\tau=\tau_0$ through Eq.~\eqref{eq:NonRes:Conditions For Gate Times} related to the $\text{CZ}_-^\text{nres}$ gate. At that gate time the fidelity $F_\text{N}$ for the respective locally equivalent CZ gate for different noise levels can be calculated as depicted in blue in figure~\ref{fig:FidNoise}. For reference, the case of a constant exchange at the first gate time $\tau_0$ (with $J_1=0$) is also shown.

The exponential decline in $\sigma$ is predicted by Eq.~\eqref{eq:NoiseFidelityAtGateTimes}. In this case, however, the gate times of the two gates $\text{CZ}_-^\text{nres}$ and $\text{CZ}^\text{const}$ are approximately the same, leading to a similar decline in $\sigma/J_0$. The fidelity of the $\text{CZ}_-^\text{nres}$ gate at the first gate time with an oscillating exchange is slightly below the fidelity with a constant exchange (see figure~\ref{Fig:NonResFid-tdep}).

\section{Conclusions \label{sec:conclusion}}
We provided a detailed theoretical description of two-qubit gates using oscillating exchange interaction between two electron spins. 
For the resonant case  with $\omega = \Delta B$ we have found conditions for parameters $(J_0, J_1, \tau)$ to obtain a locally equivalent CZ and iSWAP gate by calculating the Makhlin invariants. In fact, we have found the fidelity for these gates to be comparable to the conventional CZ gate with constant exchange. Analogously, we have calculated the numerical fidelity for the far-detuned ($|2 \Delta B -\omega| \gg J_1$) CZ gate, which performs similar as the constant exchange case. We point out that the gates found in this work can be transformed into regular CZ and iSWAP gates, respectively, by  applying $z$ rotations on the two qubits.

Moreover, we also took into account real experimental conditions by adding charge noise in terms of Gaussian distributed deviations from the static part of the exchange interaction $J_0$. We found that for relatively small charge noise contributions the fidelity of the resonant solutions for the iSWAP and CZ gates compare to the fidelity of the state-of-the-art exchange gate ($J_1=0$), but drop rather quickly for larger deviations due to longer gate times. Furthermore, the non-resonant solution lies slightly below the static case. 

Here we did not take into account any further effects impacting the fidelity. However, in real spin qubit devices the dynamical displacement of the electron in the magnetic gradient field induced by the oscillating barrier gate voltage as well as crosstalk effects due to residual driving of neighboring gate electrodes can lead to gate infidelities and need to be taken into account \cite{Heinz2021, PhysRevB.105.085414}. These effects can, e.g., be compensated using virtual gates \cite{PhysRevB.105.L121402, Hanson_2007, Volk_2019}. 

In summary, oscillating exchange gates give rise to an additional two-qubit gate in the toolbox of quantum gates for spin qubits and can perform equally well as their static counterpart when charge noise is relatively low.

\ack
This work has been supported by QLSI with funding from the European Union's Horizon 2020 research and innovation program under grant agreement No 951852 and by the Deutsche Forschungsgemeinschaft (DFG, German Research Foundation) Grant No. SFB 1432 - Project-ID 425217212.

\section*{References}
\bibliography{iopart-num}

\appendix



\end{document}